\documentclass[sigconf,screen]{acmart}

\usepackage{amsmath,amsfonts,bm}
\usepackage{booktabs}
\usepackage{array}
\usepackage{pifont}
\newcommand{\cmark}{\textcolor{black}{\ding{51}}}   
\newcommand{\xmark}{\textcolor{red}{\ding{55}}}    
\newcolumntype{C}[1]{>{\centering\arraybackslash}m{#1}}

\usepackage{graphicx}
\usepackage{subcaption}
\usepackage{wrapfig}
\usepackage{enumitem}
\usepackage{xspace}
\usepackage{url}
\usepackage{hyperref}
\hypersetup{
    colorlinks=true,
    linkcolor=blue,
    filecolor=magenta,      
    urlcolor=blue,
    pdftitle={Overleaf Example},
    }

\usepackage{microtype}
\usepackage{multirow}
\usepackage{multicol}
\usepackage{tabularx}
\usepackage{makecell}
\usepackage{tcolorbox}

\newcommand{\dataname}{CrediBench\xspace}

\newcommand{\weak}{DomainRel\xspace} 

\newtheorem{definition}{Definition}

\newcommand{\mat}[1]{\ensuremath{\mathbf{#1}}}

\newcommand{\first}[1]{\textbf{\textcolor{red}{#1}}}
\newcommand{\second}[1]{\underline{\textcolor{blue}{#1}}}

\AtBeginDocument{%
  }

\copyrightyear{2026}
\acmYear{2026}
\setcopyright{cc}
\setcctype{by}
\acmConference[KDD '26]{Proceedings of the 32nd ACM SIGKDD Conference on Knowledge Discovery and Data Mining V.2}{August 09--13, 2026}{Jeju Island, Republic of Korea}
\acmBooktitle{Proceedings of the 32nd ACM SIGKDD Conference on Knowledge Discovery and Data Mining V.2 (KDD '26), August 09--13, 2026, Jeju Island, Republic of Korea}
\acmDOI{10.1145/3770855.3817476}
\acmISBN{979-8-4007-2259-2/2026/08}

\usepackage[table]{xcolor}

\begin{document}
\title[CrediBench]{CrediBench: Web-Scale Network Data for Credibility Prediction}


\author{Emma Kondrup}
\authornote{Equal contributions.}
\affiliation{%
  \institution{McGill University}
  \institution{Mila - Quebec AI Institute}
  \city{Montreal}
  \state{Quebec}
  \country{Canada}}
\email{emma.kondrup@mila.quebec}

\author{Sebastian Sabry}
\authornotemark[1]
\affiliation{%
  \institution{McGill University}
  \institution{Mila - Quebec AI Institute}
  \city{Montreal}
  \state{Quebec}
  \country{Canada}}
\email{sebastian.sabry@mail.mcgill.ca}

\author{Hussein Abdallah}
\affiliation{%
 \institution{McGill University}
  \institution{Mila - Quebec AI Institute}
  \city{Montreal}
  \state{Quebec}
  \country{Canada}}
\email{hussein.abdallah@mila.quebec}

\author{Zachary Yang}
\affiliation{%
  \institution{Ubisoft - La Forge}
  \institution{Mila - Quebec AI Institute}
  \city{Montreal}
  \state{Quebec}
  \country{Canada}
}

\author{Jiaqi Xiong}
\affiliation{%
  \institution{University of Oxford}
  \city{Oxford}
  \country{UK}}

  \author{Kellin Pelrine}
\affiliation{%
  \institution{Mila - Quebec AI Institute}
  \city{Montreal}
  \state{Quebec}
  \country{Canada}
}

\author{James Zhou}
\affiliation{%
 \institution{University of California, Berkeley}
 \city{Berkely}
 \state{California}
 \country{USA}}

\author{Zhijin Guo}
\affiliation{
    \institution{University of Oxford}
  \city{Oxford}
  \country{UK}}

\author{Michael M. Bronstein}
\affiliation{%
  \institution{University of Oxford}
  \institution{AITHYRA Research Institute}
  \city{Oxford}
  \country{UK}}

\author{Jean-Fran\c{c}ois Godbout}
\affiliation{  \institution{Université de Montréal}
  \institution{Mila - Quebec AI Institute}
  \city{Montreal}
  \state{Quebec}
  \country{Canada}}

\author{Reihaneh Rabbany}
\affiliation{  \institution{McGill University}
  \institution{Mila - Quebec AI Institute}
  \city{Montreal}
  \state{Quebec}
  \country{Canada}}

\author{Shenyang Huang}
\affiliation{%
  \institution{University of Oxford}
  \city{Oxford}
  \country{UK}}
\email{andy.huang@cs.ox.ac.uk}
\renewcommand{\shortauthors}{Kondrup et al.}

\begin{abstract}
Automatically assessing the credibility of online sources presents an invaluable tool for navigating today’s information ecosystem. However, existing approaches either depend on scarce and costly human annotations, or focus exclusively on assessments at the level of individual claims. Misinformation often spreads via interlinked web domains, whose connections evolve over time. Focusing on claims alone ignores these structural and temporal credibility signals evident in the changing web topology. Existing datasets fail to capture these central modalities in web domain credibility prediction: namely, \emph{internet topology}, \emph{temporality} and \emph{text (webpage) content}.  To address this gap, we present \dataname, a dataset containing eights months of web graph data; of which we analyze the three months surrounding the 2024 U.S. federal elections, a time of heightened misinformation propagation online. Each monthly snapshot contains over 40 million nodes, their scraped webpage content, and over 1 billion hyperlink edges. \dataname supports credibility prediction as both a regression (continuous credibility score) and a binary classification task (credible or not). 
For classification, we curate a novel binary label set containing 662,575 web domains labelled for boolean credibility, spanning four areas (misinformation, crowd-sourced, malware and phishing). 
Our empirical experiments support that all task modalities---graph, text and time---contribute significantly to achieving the best performance. In particular, our multi-modal regression model trained on \dataname outperforms other configurations and existing baselines, decreasing Mean Average Error from \emph{0.162} to \emph{0.107} on the regression task, while the multi-modal classifier improves accuracy from 56\% to 85\% on the classification one. \textit{\dataname}, our proposed web-scale multi-modal dataset, is available on \href{https://huggingface.co/datasets/credi-net/CrediBench}{Huggingface}, providing a strong foundation for future research on credibility prediction. 
\end{abstract}





\maketitle

\section{Introduction}

 \begin{figure}[!t]
        \centering 
        \includegraphics[width=0.48\textwidth]{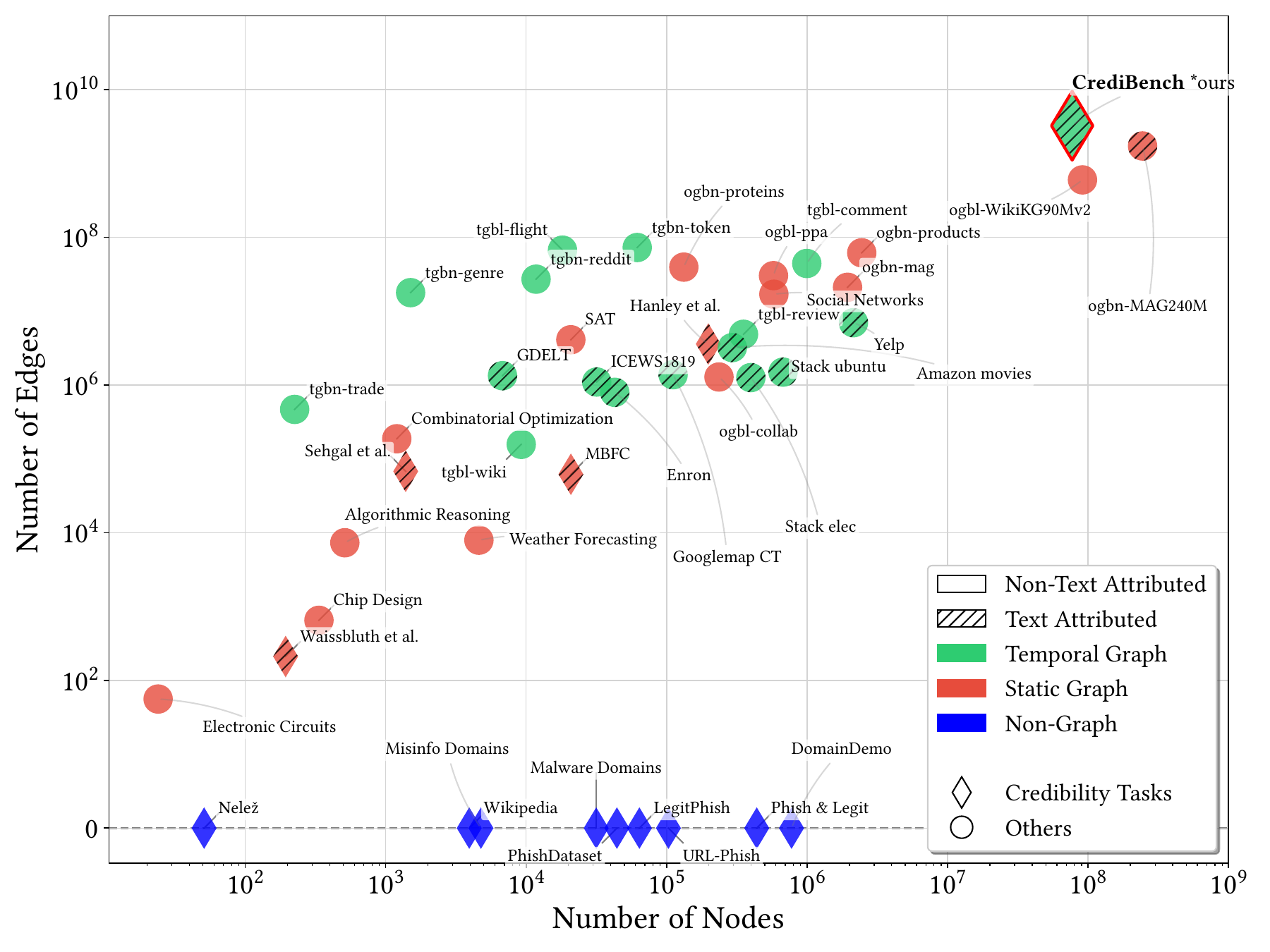}
        \caption{
        \dataname is orders of magnitude larger than most existing graph benchmarks as well as credibility datasets. \dataname is the only credibility prediction dataset that includes all three modalities: internet topology, temporality, and web text content.}
        
        \label{fig:related-work-scatter}
\end{figure}


 \begin{figure*}[t]
        \centering \includegraphics[width=0.85\textwidth]{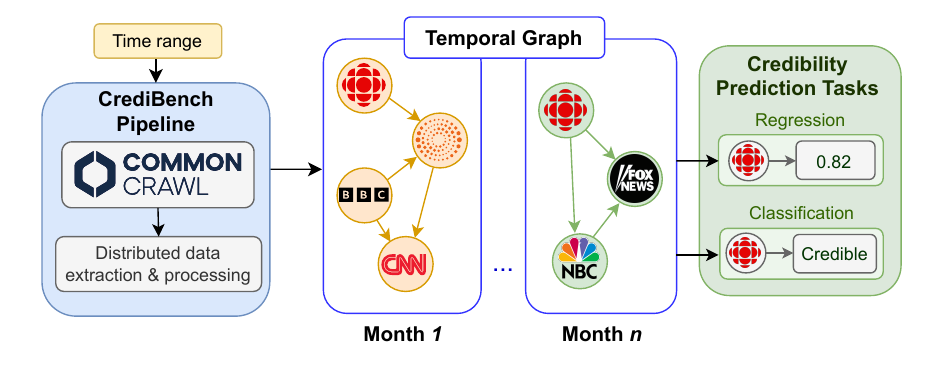}
         \caption{
         Overview of our data curation pipeline. Monthly Common Crawl snapshots are processed to construct monthly web graphs, where nodes represent web domains, edges denote hyperlinks and page content as text attributes. The resulting graphs and training labels are used to train ML models for credibility prediction tasks.
        }

        \label{fig:high-level-pipeline}
\end{figure*}

Information dissemination is increasingly occurring through web platforms, rather than traditional media outlets such as newspapers, radio or television~\cite{benkler2006wealth}. Several transformative shifts have given rise to a complex and highly dynamic information ecosystem, shaped by evolving social communities and technological infrastructures \cite{benkler2006wealth}. Notably, the open web and algorithmic social media platforms have amplified the speed and ease of access to data, which has fundamentally reshaped how information is produced, disseminated, and consumed at scale \cite{DBLP:journals/eswa/MeelV20, berinsky2023political}. Today, generative AI represents a further pivotal force within this already fragile ecosystem \cite{jaidka2025misinformation}: Large Language Models (LLMs) can generate text that is increasingly indistinguishable from human-generated content \cite{DBLP:conf/iclr/ChenS24} and persuasive \cite{LLM_Persuasiveness}; bearing political consequences of its own~\cite{costello2026large}. 
While these same conditions facilitate access to knowledge, they can also enable the rapid spread of misinformation, amplifying the reach of unreliable or misleading content, and thus its potential effect on public opinions~\cite{berinsky2023political}. The societal consequences of misinformation are now well documented: persistent exposure to unreliable information  poses concrete risks to democratic processes, public health, and social cohesion~\cite{eysenbauch_infodemiology_2002, DBLP:journals/eswa/MeelV20}. Given the scalability of AI-driven information production and dissemination, these developments underscore how largely uncontrolled the online information environment has become, and highlight the urgent need for principled, scalable mitigation strategies~\cite{jaidka2025misinformation}.

    
Source credibility assessments serve as valuable tools for navigating increasingly complex online information ecosystems \cite{Effectiveness_CRAAP_test, sultan2024}. However, they are hard to produce at scale; the concept of source credibility itself is far from trivial. Substantive work has been done around this and adjacent concepts, such as source reliability or perceived trust~\cite{whitehead1968factors}. Here, we follow the formulation of Berlo, Lemert, and Mertz~\cite{berlo1969dimensions}, which deconstructs source credibility as being jointly determined by the source's \textit{safety} (or trustworthiness), its \textit{qualification} (or competence level), and \textit{dynamism} (activity level). Our approach models patterns pertaining to each of these dimensions, as an avenue for robust credibility modelling. 

\paragraph{\textbf{Dataset Modalities.}}
Existing Machine Learning (ML) assessment methods for measuring source credibility are limited in size and scope, operating on small datasets obtained from a single platform in most cases~\cite{DBLP:conf/socialcom/ODonovanKMHA12, DBLP:journals/asc/PakaBKSC21, DBLP:journals/eswa/KumariE21, DBLP:conf/socinfo/0003KCM14, DBLP:conf/acl/Wang17, DBLP:journals/corr/abs-2501-03290}. To date, jointly leveraging all three central modalities in web domain credibility prediction (internet topology, temporal evolution and webpage text content) remains largely understudied. 
To combat the spread of misinformation, there is an urgent need to develop such methods that are deployable at the web scale. Capturing the various, dynamically-interacting signals that web entities exhibit, is not only required for many downstream use cases; but simultaneously presents an avenue for generalizability to emerging content generation methods and patterns. Existing automated approaches largely fall short of these goals~\cite{hamed_review, DBLP:conf/coinco/SupremVP22}. 
        

\paragraph{\textbf{Dataset Size.}}
A key bottleneck to the development of such methods is the sparsity of publicly available large and high-quality datasets. Indeed, existing web datasets for this task at the domain level are rare, limited in scale, and scope; a visual representation can be found in Figure~\ref{fig:related-work-scatter}. Similarly, Graph ML, and particularly Temporal Graph Learning, are limited by a lack of large, high-quality graph datasets~\cite{DBLP:journals/corr/abs-2502-14546}. Frontier benchmarks (such as the Temporal Graph Benchmark~\cite{DBLP:conf/nips/GastingerHGLPPD24}, GraphBench~\cite{DBLP:journals/corr/abs-2512-04475} or the Open Graph Benchmark~\cite{hu2021ogblsc}) primarily consist of datasets with few hundred thousands of nodes; otherwise, they mostly either lack multi-modality, temporality, or are synthetic; with an overarching lack of real-world graph benchmarks~\cite{DBLP:journals/corr/abs-2502-14546}. More discussion around existing datasets are included in Related work Section.


To accelerate the development of ML methods for domain credibility prediction, we present \dataname, a large-scale dataset containing billions of edges and tens of millions of nodes, capturing the topology of web domains, their temporal evolution, and their webpage content. Table~\ref{tab:dataset-stats} shows the dataset statistics and scale of \dataname. The Monthly snapshots are curated using our optimized web data extraction pipeline visualized in Figure~\ref{fig:high-level-pipeline}, which jointly extracts web graphs and text and efficiently processes cumulative months. On these graphs, we model credibility prediction under two task settings; \textit{regression} and \textit{binary classification}. Notably, we present an original, large and cross-domain binary classification label set of 662,575 web domains. Under these task settings, we conduct empirical studies on \dataname with text, graph and multi-modal ML methods to demonstrate the importance of combining different modalities for credibility prediction. Overall, our contributions are summarized as follows: 

\begin{table}[t]
\caption{Statistics of the \dataname monthly snapshots (October--December 2024).}
\label{tab:dataset-stats}
\centering
\small
\resizebox{1.0\columnwidth}{!}{
\begin{tabular}{lcccccc}
\toprule
\textbf{Snapshot} &
\textbf{Domains} &
\textbf{Links} &
\textbf{Text} &
\multicolumn{2}{c}{\textbf{Degree}} &
\textbf{Density} \\

\cmidrule(lr){5-6}

& & & &
\textbf{Mean} &
\textbf{Max} &
\\

\midrule
\textbf{CB-Oct24}
& 50.3M
& 1.1B
& 159.4 GB
& 42.75
& 17.1M
& $4.3\times10^{-7}$ \\

\textbf{CB-Nov24}
& 50.7M
& 1.2B
& 166.5 GB
& 45.95
& 17.3M
& $4.5\times10^{-7}$ \\

\textbf{CB-Dec24}
& 45.0M
& 1.0B
& 116.4 GB
& 45.06
& 14.7M
& $5.0\times10^{-7}$ \\

\bottomrule
\end{tabular}
}
\end{table}



\begin{itemize}[leftmargin=*]
    \item \textbf{Largest domain credibility dataset to date.} We introduce the \dataname dataset, consisting of eight months of web-scale graphs with rich text attributes. Each month contains over 40 million web domains as nodes, 1 billion of hyperlink interactions as edges, and over 100GB of the webpage content as text files. To the best of our knowledge, this is the largest domain credibility dataset as of this writing. 

    \item \textbf{Two task formulations with curated datasets.} In this work, we formulate the credibility prediction problem both as a regression task and a binary classification task. Given the sparsity of manual labels, we curate a novel dataset entitled DomainRel, the largest binary labelled set for credibility prediction, totalling 662k domains spanning across four distinct categories: misinformation, crowd-sourced, malware, and phishing. 
    
    
    \item \textbf{Extensive results demonstrating multi-modal credibility learning.} While prior work mostly focuses on a single modality for credibility prediction, our extensive experiments on \dataname demonstrate that the best performance is achieved by utilizing all three modalities for the task: graph, text and temporality. When compared to existing baselines, credibility prediction models utilizing the multi-modality of \dataname reduces MAE from \emph{0.162} to \emph{0.107} on the regression task while improving accuracy from 56\% to 85\% on the classification task.
    

\end{itemize}

\textbf{Reproducibility.} The \dataname datasets are hosted on Huggingface at \href{https://huggingface.co/datasets/credi-net/CrediBench}{CrediBench},\footnote{\url{https://huggingface.co/datasets/credi-net/CrediBench}} either downloadable there or queriable through our CrediGraph API Package (\texttt{pip install credigraph}).\footnote{Full documentation available at \url{https://credinet.readthedocs.io/en/latest/}.} 
Our curated \href{https://huggingface.co/datasets/credi-net/DomainRel}{DomainRel}, the binary labeled dataset, is also available on Huggingface.\footnote{\url{https://huggingface.co/datasets/credi-net/DomainRel}} For reproducibility and implementation, all code is publicly available under the CrediNet organisation's \href{https://github.com/credi-net}{GitHub}.\footnote{\url{https://github.com/credi-net}.}






\section{Related work}

\paragraph{\textbf{Online information ecosystems.}}
Substantial work has now established the nature of digital ecosystems as complex, socio-technical environments; shaped by social interactions~\cite{boyd2012networked}, platform and algorithm design~\cite{guess2023algorithms, guess2023reshares}, and cognitive biases~\cite{pennycook2021psychology}. Social media platforms amplify engagement-driven dynamics, which can accelerate information diffusion regardless of veracity and thereby foster homophilic echo chambers~\cite{cinelli2020echo, wang2024linguistic}. Psychological studies further demonstrate that repeated exposure~\cite{taber2009motivated}, source familiarity, and social endorsement~\cite{guess2023reshares} strongly influence perceived credibility~\cite{hasher1977frequency, taber2009motivated}. These dynamics allow misinformation to persist and compound over time, often spreading across multiple sources and platforms. The rise of generative AI also amplifies and introduces new risks, enabling the accessible and scalable production of increasingly persuasive and sophisticated content, lowering barriers for misinformation spread~\cite{jaidka2025misinformation}. Fortunately, AI also stands as an important avenue for emerging mitigation strategies. LLMs are increasingly explored for misinformation detection \cite{DBLP:journals/aim/ChenS24, DBLP:conf/emnlp/PelrineITRGCGR23, jiang_disinformation_2023}. Particularly, Retrieval-Augmented Generation (RAG) can aid in evidence-based misinformation detection, as introduced in  \cite{DBLP:journals/corr/abs-2409-00009}.


\paragraph{\textbf{Methods for source assessments.}}
Efforts to assess the credibility of online information sources have ranged from manual, user-~\cite{CRAAP, SIFT} or expert-driven \cite{dataCredibility, lasser_misinformation_domains} evaluations to computational methods. 
Most web credibility datasets rely on advanced expert assessments for label annotation \cite{dataCredibility, lasser_misinformation_domains}. Such manual methods are inherently unable to scale to modern information flows. To address this, machine and deep learning methods quickly emerged as a strong solution. As highlighted in \cite{DBLP:journals/corr/abs-2411-05060}, such efforts have almost exclusively focused on classifying individual claims, posts, or documents. These usually rely on endogenous signals from the text, such as linguistic and stylistic features, semantic content, and multimodal cues \cite{DBLP:conf/icde/ZhangDY26, DBLP:journals/eswa/ChoudharyA21, DBLP:journals/mta/KaliyarGN21, DBLP:conf/cikm/RuchanskySL17, DBLP:journals/eswa/KumariE21}. 



\paragraph{\textbf{Structural and temporal components of information dissemination.}}
In practice, information is not consumed in isolation; rather, it exists in contextual and interactive environments where information is shared and endorsed~\cite{sultan2024}. Misinformation thus spreads across sources and platforms. Detecting whether a certain piece of text contains misinformation, largely benefits from assessing its source, its output patterns, and interactions with other web entities \cite{vosoughi2018spread, shao2018spread}. It has indeed been demonstrated, that showing the source with a piece of information, increases misinformation detection capabilities and can thus help lower its spread \cite{sultan2024}. Furthermore, the structural dimensions that play a key role in misinformation dissemination~\cite{daron_acemoglu_model_2022, vosoughi2018spread}, motivate the need for methods that move past individual, endogenous assessments. Exogenous signals, such as social endorsement cues, have thus also been integrated in some works \cite{DBLP:conf/www/CastilloMP11, DBLP:journals/asc/PakaBKSC21, DBLP:conf/socialcom/ODonovanKMHA12}, but remain heavily limited. They vastly operate on small-scale social graphs or a single-platform scope \cite{DBLP:conf/www/CastilloMP11, DBLP:conf/socinfo/0003KCM14}, usually static or spanning only a short time window \cite{xu2024offline}. This can significantly limit their generalizability~\cite{sultan2024}. Implicitly integrating complex patterns as such, if at all, remains insufficient for many applications \cite{xu2024offline, hoy2025generalisability}. Global and reliable assessments at web-scale require principled methods that are larger in scope and incorporate the different factors that can help determine whether to rely on a source; including its behavioral and interactional patterns. 


\paragraph{\textbf{Source Credibility Datasets.}}
The lack of labelled data, and of quality datasets at large, stands as a bottleneck both for misinformation research, and for fields of Graph ML. This stands particularly true of domain-level datasets for credibility tasks, which often stand at the intersection of the two, and are hard to produce given their scale and nature. Indeed, while the value of source assessments is well established, a vast majority of existing misinformation datasets are at the claim level \cite{DBLP:journals/corr/abs-2411-05060, DBLP:conf/acl/Wang17}. Figure~\ref{fig:related-work-scatter} shows an ensemble of existing datasets, their methodology, modality and scope; allowing us to compare not only existing misinformation datasets, but graph datasets in general \cite{DBLP:conf/nips/GastingerHGLPPD24, hu2021ogblsc, DBLP:journals/corr/abs-2512-04475, NEURIPS2024_a65d054a}. As it shows, the scale and richness of existing datasets is limited. The sensibility of certain aspects of web data also poses limitations; notably, multiple of these datasets are limited by the fact they are not publicly released \cite{DBLP:journals/corr/abs-2104-11694, DBLP:journals/corr/abs-2205-03338, DBLP:conf/icwsm/Hanley0D22}. 


Among the limited lines of work that focus on the domain level, there are some heterogeneous datasets that link websites to social media context such as \cite{williams2024bridging} or DomainDemo \cite{DBLP:journals/corr/abs-2501-09035}.  \cite{DBLP:conf/icwsm/Hanley0D22} and \cite{DBLP:journals/pacmhci/Hanley0D23} are two similar works that built web graph datasets restricted to websites linked to the far-right QAnon conspiracy theories. \cite{DBLP:journals/corr/abs-2104-11694} and \cite{DBLP:journals/corr/abs-2205-03338} built domain-level hyperlink networks between a few thousand known misinformation domains, and a similar number of informational ones. The MBFC* dataset~\cite{DBLP:conf/icwsm/CarragherWC24} is another web graph dataset based on a seed set from MBFC~\cite{MBFC} and a few other fact-checking resources, yet it has less than 1 million edges, and is still among the largest publicly available dataset of this nature.


\paragraph{\textbf{Graph Representation Learning.}}
Graph modelling provides a natural framework for source credibility, presenting methods that can intrinsically combine content, structural and temporal patterns. Graphs enable the representation of complex relationships between sources and the propagation of credibility signals~\cite{moorthy_dual_2025}. Early approaches leveraged link analysis algorithms~\cite{DBLP:journals/cn/BrinP98}, using them as features for learning~\cite{DBLP:conf/ecir/OlteanuPLA13}, or exploiting graph knowledge bases~\cite{hu_compare_2021}. Most use modern Graph Neural Networks (GNNs) to capture propagation patterns of credibility signals through message passing \cite{DBLP:journals/corr/TacchiniBVMA17, DBLP:journals/tjs/ZhaoZR24, DBLP:conf/icdm/RenWZ020, DBLP:journals/corr/abs-2501-03290}. 
For instance, \cite{DBLP:journals/tjs/ZhaoZR24} propose a dual-channel graph attention mechanism that jointly captures link structure and propagation dynamics, and \cite{DBLP:conf/icdm/RenWZ020} integrates heterogeneous graph data with adversarial active learning to handle noise and diversity. However, these efforts are limited by a lack of real-world, high-quality datasets, especially at scale. Frontier benchmark suites, which have underpinned much of the area's research efforts over recent years (such as the Open Graph Benchmark (OGB)~\cite{hu2021ogblsc} or GraphBench~\cite{DBLP:journals/corr/abs-2512-04475}), contain datasets that all incur some of the fundamental limitations for the aforementioned applications. Many lack temporality, consist of synthetic data~\cite{DBLP:journals/corr/abs-2512-04475} or are scoped on a single platform~\cite{DBLP:conf/nips/GastingerHGLPPD24, hu2021ogblsc, DBLP:journals/corr/abs-2512-04475}, and the suites vastly consist of datasets in the low hundreds of thousands of nodes. While OGB focuses on making available large-scale graph datasets, including temporal, billion-scale ones~\cite{hu2021ogblsc}, these are still importantly limited in scope.~\footnote{E.g., OGB's largest temporal dataset, \texttt{ogbn-papers100M}~\cite{hu2021ogblsc}, is a citation graph with about 1.6B interactions, presenting a large, real-world graph that is yet not general-scoped.}

\section{Preliminaries} \label{sec:prelim}

Our graph construction pipeline produces Text-Attributed temporal Graphs (TAGs) for web interaction credibility modelling. We then establish and study credibility prediction as a TAG learning problem under two task settings: regression (predicting continuous scores) and classification (predicting a boolean value).

\subsection{Text-Attributed temporal Graphs (TAG)} 
We first introduce the notation for our temporal text-attributed graphs. As the Common Crawl data is released monthly, we represent them as sequences of graph snapshots, similar to the Discrete-time Dynamic Graph setup defined in~\cite{DBLP:journals/jmlr/KazemiGJKSFP20}. The data formulation is as follows: 

\begin{definition}[Text-Attributed temporal Graphs] 
A Text-Attributed temporal Graph (TAG) $\mat{G}$ is a sequence of graph snapshots sampled at regularly-spaced time intervals and nodes have evolving text content at each snapshot:
\begin{equation*}
\mat{G} = \{ \mat{G}_0, \mat{G}_1, \dots, \mat{G}_T \}
\end{equation*} where $\mat{G}_t = \{\mat{V}_t, \mat{E}_t\, \mat{X}_t$ \} is the graph at timestamp $t\in [0,T]$, where $\mat{V}_t \in \mathbb{R}^{|\mat{V}_t|}$, $\mat{E}_t \in \mathbb{R}^{|\mat{V}_t| \times |\mat{V}_t|}$ are the set of nodes and edges in $\mat{G}_t$ and $\mat{X}_t \in \mathbb{R}^{|\mat{V}_t| \times L}$ is the text feature of nodes in $\mat{G}_t$ where $L$ is the maximum text length. Note that $\mat{X}_t$ stores the raw text content from each web domain and can subsequently be processed into text embedding vectors for models to use. 
\end{definition}

\subsection{Credibility Prediction}
We model credibility prediction as a node-level prediction task, as follows. 


\begin{definition}[Node Prediction Task] 

Let $\mat{G}_t=(\mat{V}_t,\mat{E}_t,\mat{X}_t)$ be snapshot $t$ of TAG $\mat{G}$. Let $y_t:\mat{V}_t\rightarrow\mathcal{Y}$ be a (possibly time-varying) node labeling function.
The temporal node prediction task is to learn a function
\begin{equation}
f_t:\mat{V}_t \times \mat{G}_{\le t} \rightarrow \mathcal{Y}
\end{equation}
that predicts $y_t(v)$ for all $v\in\mat{V}_t$, using graph structure and node attributes up to time $t$. In Regression, $\mathcal{Y}$ operates on range $[0,1]$, whereas for Binary Classification its range is $\{0,1\}$. 
\end{definition}









\begin{table}[t]
\centering
\caption{We model credibility prediction on \dataname graphs, both as a regression task (Reg.) and a binary classification task (Cls.) for credibility prediction. The binary labelled set is curated for this task from eight existing datasets.} \label{tab:labels_breakdown}
\setlength{\tabcolsep}{5pt}
\renewcommand{\arraystretch}{0.85}
\resizebox{\columnwidth}{!}{
\begin{tabular}{llcr}
\toprule
\textbf{Task} & \textbf{Domain} & \textbf{Label Distribution} & \textbf{Total} \\
\midrule 
& \footnotesize{Crowd-sourced} & \raisebox{-0.45\height}{\includegraphics[width=0.06\textwidth,trim=0 0 0 5cm,
    clip]{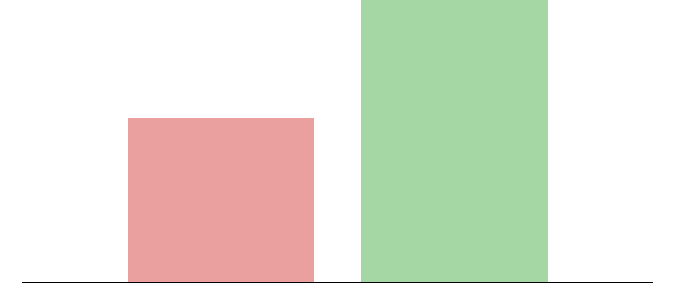}} & 3{,}935 \\
\cmidrule{2-4}
& \footnotesize{Misinformation} & \raisebox{-0.45\height}{\includegraphics[width=0.06\textwidth,trim=0 0 0 5cm,
    clip]{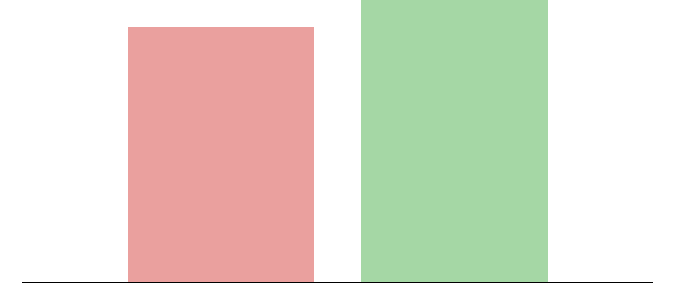}} & 4{,}818 \\
\cmidrule{2-4}
& \footnotesize{Phishing} & \raisebox{-0.45\height}{\includegraphics[width=0.06\textwidth,trim=0 0 0 5cm,
    clip]{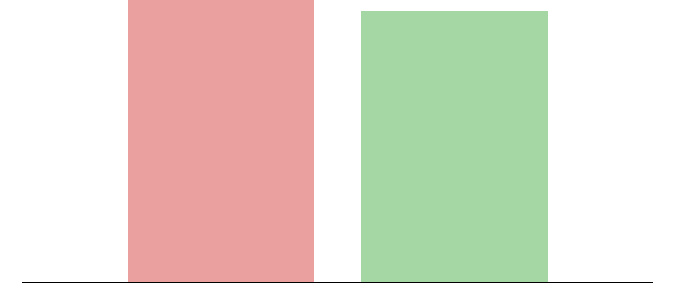}} & 649{,}825 \\
\cmidrule{2-4}
& \footnotesize{Malware} & \raisebox{-0.45\height}{\includegraphics[width=0.06\textwidth,trim=0 0 0 5cm,
    clip]{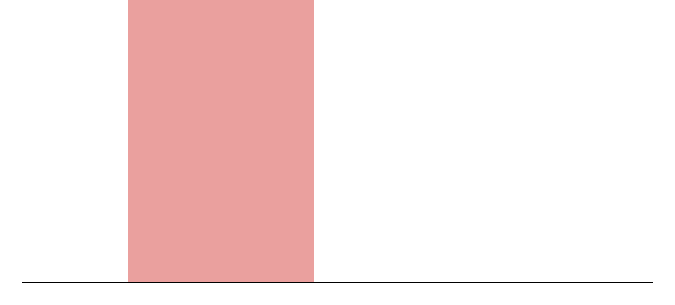}} & 31{,}540 \\
\cmidrule{2-4}
\rowcolor{gray!10}
 Class. & DomainRel & \raisebox{-0.4\height}{\includegraphics[width=0.12\textwidth,height=1.2cm,trim=0 0 0 3cm, clip]{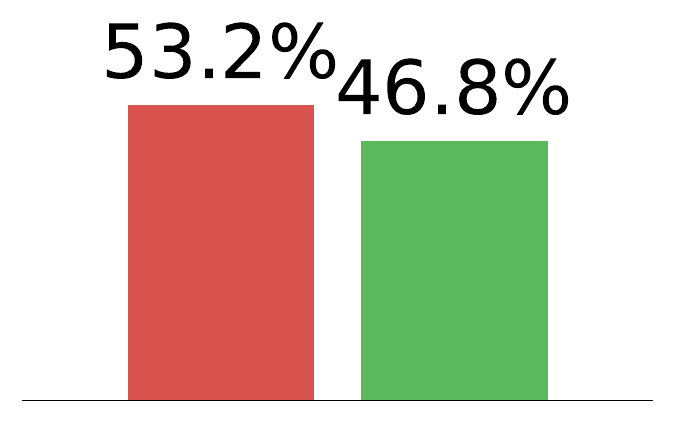}} & 662,575 \\

\midrule
\rowcolor{gray!10}
Reg.& News & \raisebox{-0.4\height}{\includegraphics[width=0.15\textwidth,height=1cm,trim=0 0 0 0.5cm, clip]{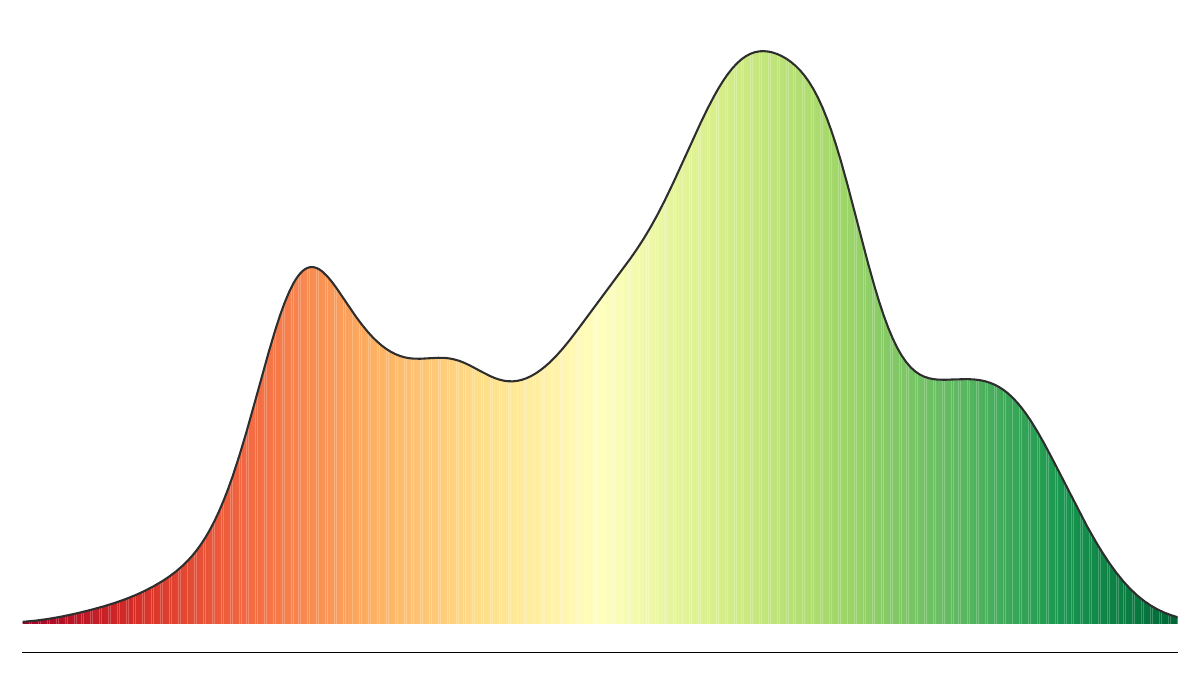}} & 11,520 \\
\bottomrule
\end{tabular}
}

\end{table}

\section{Label Sets} \label{sec:datasets}

As established, the lack of large-scale benchmark stands as a bottleneck both for graph ML, and misinformation research at large. Our approach stands as a first step towards incorporating the various fundamental factors of source and page credibility at scale, while remaining accessible. This is essential for an effective misinformation detection benchmark in the rapidly evolving modern online environment. 






\paragraph{\textbf{Regression scores.}}
The  Domain Quality Ratings (DQR)~\cite{dataCredibility} dataset is among the largest sets of domains labelled with expert-driven credibility scores, comprising 11,520 such annotated domains collected from six expert sources, including media organizations and independent professional fact-checkers. These sources apply different scoring schemes and evaluate domains across multiple dimensions (e.g., trust, reliability, bias, and transparency). 
We employ DQR's per-domain First Principal Component (PC1) score, which is derived from a Principal Component Analysis (PCA) of the six scoring sources, as the ground truth scores used for regression. This aggregate ranges from $[0,1]$. 





\paragraph{\textbf{Classification labels.}}
To address the limitation of current label sets in terms of size and domain category coverage, we introduce the \textit{Domains by Reliability} (\weak) dataset, an aggregation of over 662,575 domains normalized cross-categories and multi-lingual credibility score sources. This curated dataset is collected from eight fact-checking, public, and academic sources, covering four areas: misinformation, crowd-sourced, phishing and malware domains.  Table~\ref{tab:labels_breakdown} shows the composition breakdown. For misinformation domain, we use two expert-driven datasets (from academic and independent sources respectively)~\cite{lasser_misinformation_domains, nelez_disinformation_sites_dataset_2025}. We also use an aggregate set of cross-domain credibility ratings from multiple crowd-sourced lists from Wikipedia~\cite{kynoptic_wikipedia_reliable_sources_2026}.  We use four URL-based datasets that label websites as either being considered legitimate, or phishing websites: LegitPhish~\cite{LegitPhish}, PhishDataset~\cite{PhishDataset}, URL-Phish~\cite{damminh_urlphish_2025}, and Phish \& Legit~\cite{phish_and_legit}. While the proportion of phishing datapoints outnumbers other domains, we consider this to be at least slightly offset by their short-lived nature: despite their larger proportion overall, their coverage rates can be expected to vary considerably across different snapshots. We further supplement the phishing URLs, with Malicious and Malware domains as reported by URLhaus, a platform dedicated to sharing malicious URLs that are being used for malware distribution.
Aggregated and deduplicated, the domains garnered here form a binary labelled set of 662,575 domains; where 0 denotes an unreliable domain, and 1 its reliable counterpart. 
To the best of our knowledge, this is the largest such dataset operating at the domain level, with existing ones usually containing within the tens of thousands of domains. The labels distribution for both classification and regression datasets is shown in Table ~\ref{tab:labels_breakdown}, and the data source breakdown per category in Table \ref{tab:labels_breakdown_indetail} in the Appendix.

 \begin{figure}[htbp!]
        \centering        
        \includegraphics[width=\columnwidth]{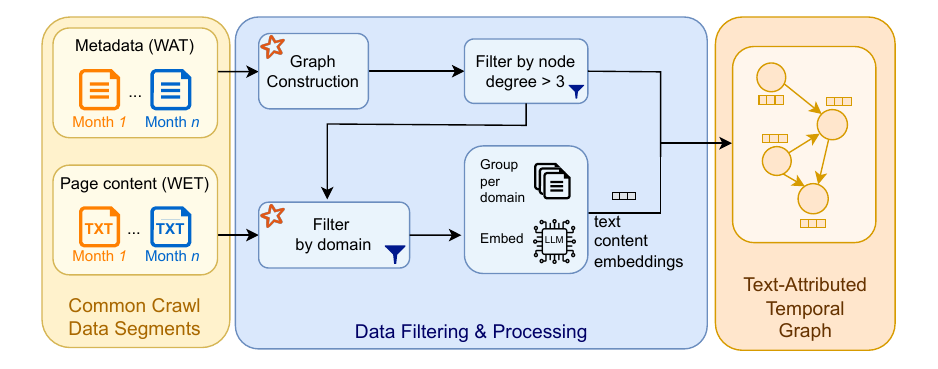}
        \caption{The graph construction pipeline extracts and processes graph and text data to produce monthly TAGs, retrieving each node’s webpage content from Common Crawl or web scraping and converting it into embeddings.}
        
        
        \label{fig:CC-graph-construction-pipeline}
\end{figure}

\section{Constructing Web-Scale TAGs}


We present an optimized and user-friendly graph construction pipeline that builds large-scale, text-attributed, temporal web graphs. Each monthly snapshot is sourced from Common Crawl \cite{Common-Crawl} and processed into a single connected graph. Figure~\ref{fig:CC-graph-construction-pipeline} shows the detailed pipeline which yields such graphs in monthly snapshots; where each snapshot contains upwards of a billion edges, which represent hyperlink interactions between domains. For more details on the pipeline's design, resource consumption, data formatting, and processing protocols enplaced, refer to Appendix~\ref{app:CG}. Accompanying our aim to enable further research in this area, we release a user-friendly PyPI Package. It makes available an easy-to-use API endpoint to query credibility scores of domains; for more details, refer to our \href{https://credinet.readthedocs.io/en/latest/}{\texttt{API documentation}}.  





\paragraph{\textbf{Graph construction.}}
The first step in graph construction is to download and decompress the raw data based on Common Crawl's decompression pipeline \cite{CC-github}. The downloaded files contain about 90,000 instances per month, which are iteratively processed in batches of 300 for computational feasibility. This batch processing is parallelized across CPUs, resulting in a graph construction spanning only a few days, compared to existing methods being considerably more resource- and time-consuming. After decompressing the files, metadata and domain hyperlinks are extracted, resulting in a web graph where each node is a web domain, and each edge a hyperlink from one web domain's page to another. 

\begin{figure}[t]
  \centering
   \begin{subfigure}[t]{0.155\textwidth}
    \centering
    \includegraphics[width=\linewidth]{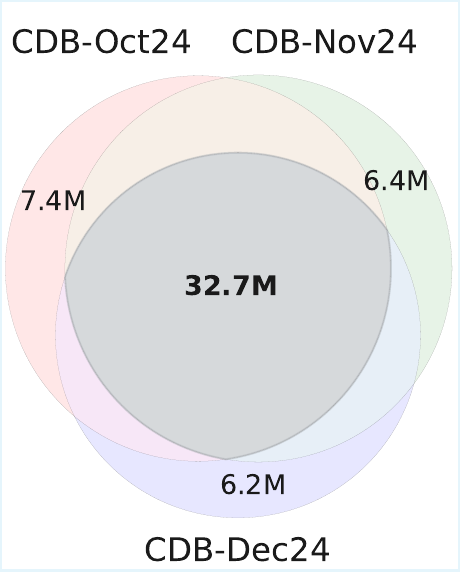}
    \caption{Domains}
    \label{fig:oct_nov_dec_domains_ven}
  \end{subfigure}
   \hfill
     \begin{subfigure}[t]{0.16\textwidth}
    \centering    
    \includegraphics[width=\linewidth]
    {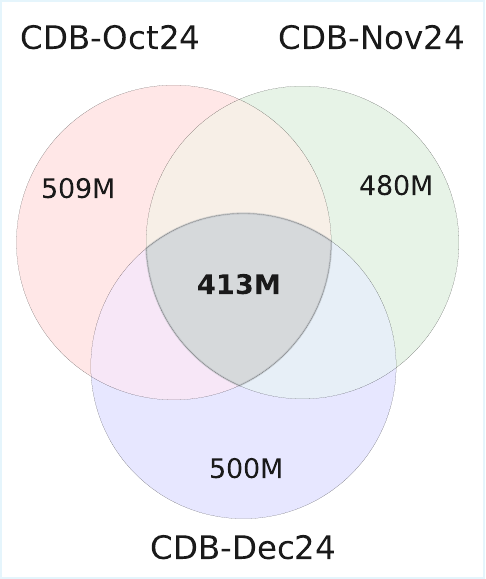}
    \caption{Graph}
    \label{fig:oct_nov_dec_edges_ven}
     \end{subfigure}
      \hfill
      \begin{subfigure}[t]{0.145\textwidth}
    \centering    
    \includegraphics[width=\linewidth]
    {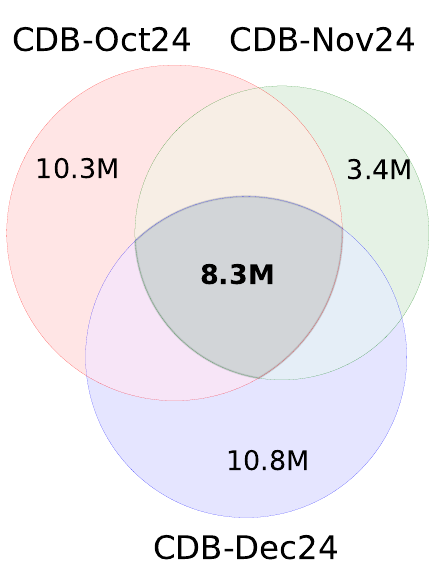}
    \caption{Content}
    \label{fig:oct_nov_dec_content_ven}
     \end{subfigure}\     
    \caption{
    Web graphs vary notably across the three snapshots: common domains account for 49\% of the total, common edges represent only 19\%, while common domains that have evolving content represent ~26 \%.  
   }
      \label{fig:domains_evolution}
\end{figure}


\paragraph{\textbf{Graph processing \& domain filtering.}}
Following the data decompression and graph construction, we attach timestamps, at the monthly granularity provided by Common Crawl. We filter domains in the graph to discard nodes with degree below a threshold, which defaults to 3, to keep a single connected component with relatively active domains. Following this degree-based filtering, we further remove any newly-isolated node. These processing steps are motivated by the fact that lower-degree (especially isolated) nodes are less likely to be relevant during retrieval queries or generally in information searches, and removing them lightens the computational load of handling the graph. 

\paragraph{\textbf{Text content extraction.}}
We extract text content for web domains present in the graph by loading their scraped content provided by the Common Crawl WET file format, which stores the extracted plain body text of pages from the original HTML. A distributed Spark cluster is used to process the monthly index and identify the WET files corresponding to the target domains, eliminating the need to download and process the full monthly WET corpus (approximately 7.3 TB \cite{cc-Dec-2024}) and reducing the workload to about 2 TB, processed in batches adjustable to the available compute resources. 
Each job extracts metadata such as text content, scraping timestamp, and content languages for the subsequent text classification task. 
The number of textual documents associated with each domain varies according to its update frequency.
In the document grouping step, all documents belonging to a given domain alongside their timestamps are appended into a single group to facilitate subsequent embedding generation. Appendix \ref{appendix:content_extract} details content filtering steps and computing requirements.

\paragraph{\textbf{The \dataname Dataset}}
To understand the structural and textual characteristics of datasets collected via our graph construction pipeline, we construct and openly release \dataname, a dataset consisting of 8 monthly snapshots.\footnote{Crawl IDs \texttt{CC-MAIN-2024-42}, \texttt{CC-MAIN-2024-46}, \texttt{CC-MAIN-2024-51}, \texttt{CC-MAIN-2025-05}, \texttt{CC-MAIN-2025-08}, \texttt{CC-MAIN-2025-13}, \texttt{CC-MAIN-2025-18} and \texttt{CC-MAIN-2025-21}.} We study a subselection of three months (September, October, November),\footnote{Crawl IDs \texttt{CC-MAIN-2024-42}, \texttt{CC-MAIN-2024-46}, and \texttt{CC-MAIN-2024-51}} 
surrounding the 2024 United States Presidential Election, a period of increased polarization and disinformation propagation~\cite{smith2024ethics}. 

\paragraph{\textbf{Domain variations.}}
Web graphs evolve over time, and their domains and connections with it. Figure~\ref{fig:domains_evolution} illustrates this through Venn diagrams of domain and graph links overlap in \dataname. In Figure~\ref{fig:domains_evolution}.a, 32.7 million domains appear in all three monthly snapshots, which represent 49\% of the overall unique domains, while October, November, and December contain 7.4, 6.4, and 6.2 million unique domains, respectively. Table \ref{tab:transition_stats} further shows the proportions of Newly Added, Deleted and Retained Months across the 3 months. The evolution of edges is also illustrated in Figure~\ref{fig:domains_evolution}.b, where only a small proportion of common edges exists across the three monthly snapshots, approximately 413M out of 2.2B unique edges that represents 19\%. The high ratio of unique edges per month underscores the potential of employing temporal GNNs to effectively capture and learn the evolving graph structure. 

\begin{table}[h]
\centering
\small
\setlength{\tabcolsep}{6pt}
\begin{tabular}{lcccccc}
\toprule
&
\multicolumn{3}{c}{\textbf{Domains}} &
\multicolumn{3}{c}{\textbf{Links}} \\
\cmidrule(lr){2-4}
\cmidrule(lr){5-7}
\textbf{Transition}
& New
& Del.
& Retained
& New
& Del.
& Retained \\
\midrule

Oct $\rightarrow$ Nov
& 19\%
& 15\%
& 80.4\%
& 52\%
& 54.7\%
& 57\% \\

Nov $\rightarrow$ Dec
& 28\%
& 24\%
& 79\%
& 54.4\%
& 59.4\%
& 47\% \\

\bottomrule
\end{tabular}
\caption{Snapshot transition statistics between consecutive temporal snapshots.
$\uparrow$ denotes newly added content,
$\downarrow$ deleted content, and
$\leftrightarrow$ retained content.}
\label{tab:transition_stats}
\end{table}
\section{Experiments}

\begin{table}[t]
{
\caption{Binary classification and regression results for the CDB-Dec24 snapshot. Classification metrics are Accuracy (\%) and F1 (macro); regression metrics are MAE and Max(AE).
Best \first{first} and \second{second} results are highlighted in bold.
A multi-modal GAT with text embeddings achieves the best overall performance.
An asterisk denotes existing works that we augmented for more insightful comparison.}
\label{tab:dec-all}
\centering
\resizebox{0.48\textwidth}{!}{%
\begin{tabular}{l l l l cc cc}
\toprule
& \multirow{2}{*}{\rotatebox[origin=c]{90}{\textbf{Graph}}}
&  \multirow{2}{*}{\rotatebox[origin=c]{90}{\textbf{Text}}}& \textbf{Method}
& \multicolumn{2}{c}{\textbf{Classification}}
& \multicolumn{2}{c}{\textbf{Regression}} \\
\cmidrule(lr){5-6}\cmidrule(lr){7-8}
& & & & \textbf{Acc.} & \textbf{F1} & \textbf{MAE} & \textbf{MAX(AE)} \\
\midrule

\multirow{5}{*}{\rotatebox[origin=c]{90}{\textbf{Baselines}}}
& \xmark & \xmark & Constant 
& 53.0 & 69.1 & 0.167 & 0.546 \\

& \cmark & \xmark &
SEO-based GNN~\cite{DBLP:conf/icwsm/CarragherWC24}
& 50.0 & 66.7 & 0.428 & 0.956 \\

& \cmark & \xmark &
LightGBM~\cite{Kadkhoda2025DomainQuality}
& 56.0 & 70.7 & 0.145 & 0.630 \\

\cmidrule(l){4-8}
& \xmark & \cmark &
LLM-URL~\cite{DBLP:conf/websci/YangM25}
& 52.84 & 63.9 & 0.162 & 0.765 \\


& \xmark & \cmark &
LLM-URL + Web*
& 54.35 & 63.03 & 0.158 & 0.719 \\

\midrule

\multirow{4}{*}{\rotatebox[origin=c]{90}{\textbf{Ours}}}
& \xmark & \cmark &
Text (OpenAI-TE3L)
& 63.2\scriptsize{$\pm$0.02}
& 60.5
& 0.118\scriptsize{$\pm$0.001}
& \first{0.506} \\

& \cmark & \xmark &
GAT (RNI)
& 68.9\scriptsize{$\pm$0.18}
& 69.7
& 0.128\scriptsize{$\pm$0.001}
& 0.763 \\

& \cmark & \cmark &
Graph $\oplus$ Text 
& \first{83.6\scriptsize{$\pm$0.03}} & \first{83.2}
& \first{0.111\scriptsize{$\pm$0.001}}
& \second{0.611} \\

& \cmark & \cmark &
GAT (Gemma3)
& \second{76.1\scriptsize{$\pm$0.36}}
& \second{75.2}
& \second{0.117\scriptsize{$\pm$0.001}}
& 0.664 \\
\bottomrule
\end{tabular}
}
}
\end{table}

To understand which aspects of the web graph data contributes to the prediction of domain credibility scores, we conduct a series of empirical experiments on the \dataname dataset. More specifically, our goal was to answer the following research questions:
\begin{itemize}[leftmargin=10pt]
    \item \textbf{RQ1:} \emph{to what extent does the text content of a domain indicates its credibility score?}
    \item \textbf{RQ2:} \emph{to what extent does the hyperlink structure of a web domain inform its credibility score?} 
    \item \textbf{RQ3:} \emph{to what extent can we benefit from combining the graph structure with the text content?} 
    \item \textbf{RQ4:} \emph{to what extent the temporal evolution of content and graph structure can benefit the credibility score prediction?}
    \item \textbf{RQ5:} \emph{how effectively does the trained binary classification model transfer to different types of domains, i.e., different heterogeneous notions of credibility?}
     \item \textbf{RQ6:} \emph{how effectively does the trained binary classification model transfer across time to future graph snapshots?}
\end{itemize}

\subsection{Experimental Setup}

\begin{figure}[t]
    \centering    \captionsetup{justification=centering}
    \begin{subfigure}[t]{0.23\textwidth}
        \centering
        \includegraphics[width=\textwidth]{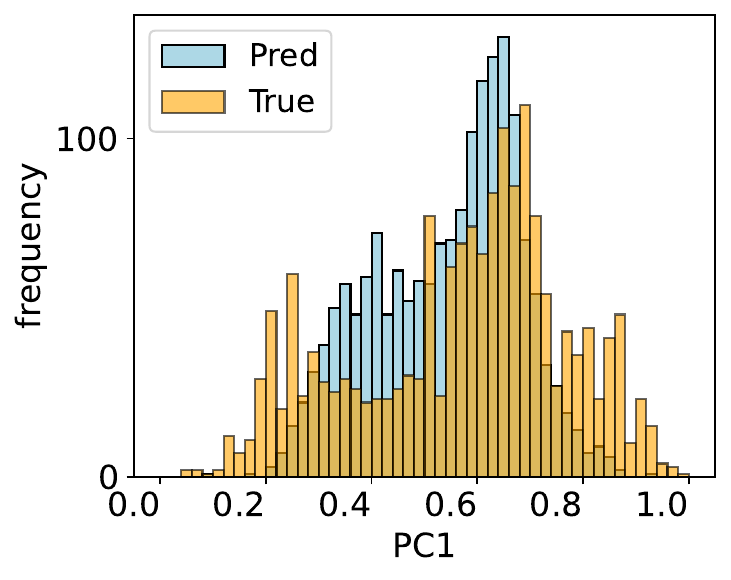}
        \caption{Text: MLP(TE3L). MAE=0.118}
        \label{fig:mlp-pc1-dist}
    \end{subfigure}
    \hfill
    \begin{subfigure}[t]{0.23\textwidth}
        \centering
        \includegraphics[width=\textwidth]{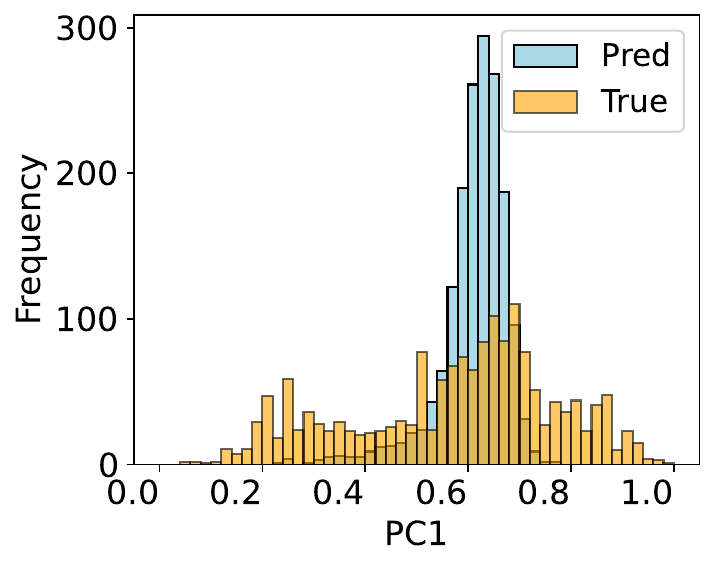}
        \caption{Graph: GAT(RNI). MAE=0.128}
        \label{fig:gat-rni-pc1-dec}
    \end{subfigure}
    \hfill
    \begin{subfigure}[t]{0.23\textwidth}
        \centering        \includegraphics[width=\textwidth]{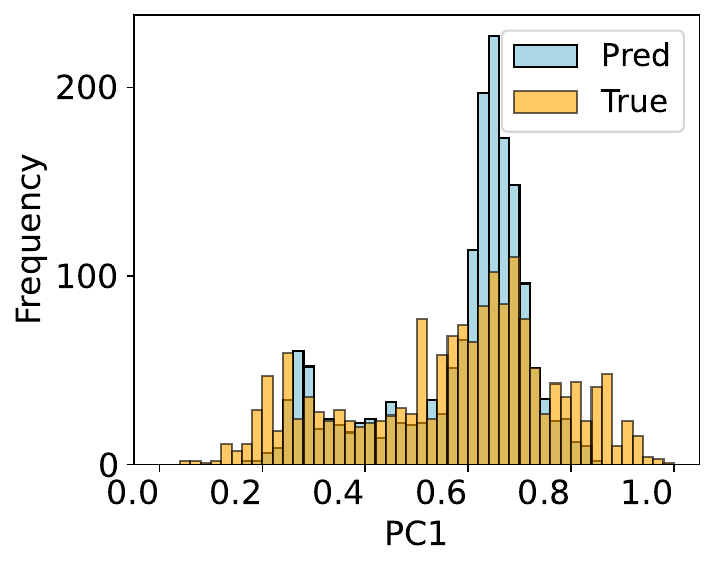}
        \caption{Graph+Text: GAT (Gemma3). MAE=0.117}
        \label{fig:gat-gemma-pc1-dec}
    \end{subfigure}
    \hfill
    \begin{subfigure}[t]{0.23\textwidth}
        \centering        \includegraphics[width=\textwidth]{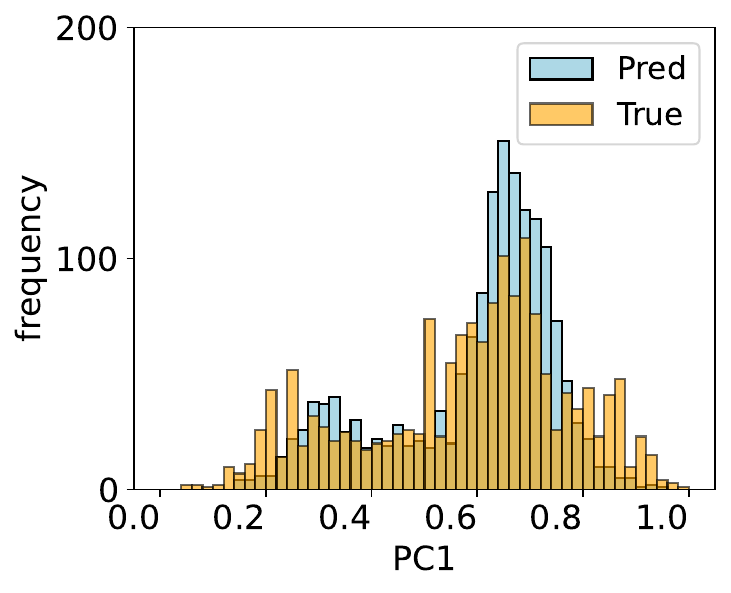}
        \caption{Temporal(Text+Graph).  MAE=0.107}
        \label{fig:mlp-pc1-dec-te3l-gat-avg}
    \end{subfigure}
    \label{fig:mlp-pc1-dec-comparison}
    \hfill
    
    \caption{PC1 score distribution for CDB-Dec24 snapshot.}
    \label{fig:regression-dist-dec-comparison}
\end{figure}

\begin{table}[t]
{
\caption{Regression credibility score (PC1) across multimodal models. Best \first{first} and \second{second} results are highlighted.}
\label{tab:regression-all}
\centering
\setlength{\tabcolsep}{1pt} %
\resizebox{0.48\textwidth}{!}{%
\begin{tabular}{l l ll ll ll}
\toprule
& \textbf{Target} & \multicolumn{2}{c}{\textbf{CDB-Oct24}} &\multicolumn{2}{c}{\textbf{CDB-Nov24}} & \multicolumn{2}{c}{\textbf{CDB-Dec24}}\\
& \textbf{Metric} & \textbf{MAE} &\textbf{AE}$_{\max}$& \textbf{MAE} &\textbf{AE}$_{\max}$ & \textbf{MAE} &\textbf{AE}$_{\max}$\\
\midrule
\multirow{5}{*}{\rotatebox[origin=c]{90}{\small \textbf{\makecell{Baselines}}}} & Mean & 0.167 & 0.546 & 0.167 & 0.546 & 0.167 & 0.546\\


& SEO-based GNN~\cite{DBLP:conf/icwsm/CarragherWC24} &  0.428 & 0.956 &  0.428 & 0.956 &  0.428 & 0.956 \\
& LightGBM~\cite{Kadkhoda2025DomainQuality} & 0.145 & 0.630 & 0.145 & 0.630 & 0.145 & 0.630 \\ 
\cmidrule(l){2-8}
& LLM-URL~\cite{DBLP:conf/websci/YangM25} & 0.162 & 0.765   & 0.162 & 0.765   & 0.162 & 0.765  \\ 
& LLM-URL~\cite{DBLP:conf/websci/YangM25} + WS* & 0.158 & 0.719 & 0.158 & 0.719 & 0.158 & 0.719\\ 
\midrule
\multirow{4}{*}{\rotatebox[origin=c]{90}{\small \textbf{Text}}} & MLP(Gemma3) &0.134\scriptsize{$\pm$0.001}&0.594&0.135 \scriptsize{$\pm$ 0.001} &0.527& 0.142 \scriptsize{$\pm$ 0.001} &0.557\\
& MLP(Qwen3-0.6B) &0.137\scriptsize{$\pm$0.001}&0.561&0.129\scriptsize{$\pm$0.001}&0.565&0.130\scriptsize{$\pm$0.001}&0.623\\
& MLP(Qwen3-8B)  &0.125\scriptsize{$\pm$0.001}&0.550&0.126\scriptsize{$\pm$0.001}&0.545&0.126\scriptsize{$\pm$0.001}&0.550\\
& MLP(TE3L)  &0.121\scriptsize{$\pm$0.001}&\second{0.541}&0.119\scriptsize{$\pm$0.001} & \first{0.506}&0.118\scriptsize{$\pm$0.001} & \first{0.506}\\
\midrule
\multirow{3}{*}{\rotatebox[origin=c]{90}{\small \textbf{\makecell{Graph}}}}& 
GCN &0.132\scriptsize{$\pm$0.005}&0.813&0.132\scriptsize{$\pm$0.001}&0.806&0.133\scriptsize{$\pm$0.001}&0.812\\
& SAGE  &0.152\scriptsize{$\pm$0.003}&0.871&0.145\scriptsize{$\pm$0.002}&0.913&0.149\scriptsize{$\pm$0.001}&0.854\\
& GAT&0.125\scriptsize{$\pm$0.002}&0.826&0.127\scriptsize{$\pm$0.001}&0.764&0.128\scriptsize{$\pm$0.001}&0.763\\
\midrule
\multirow{8}{*}{\rotatebox[origin=c]{90}{\small \textbf{\makecell{Text \& Graph}}}}& 
MLP (GAT $\oplus$ Gemma3) &0.115\scriptsize{$\pm$0.001}&0.578&0.120\scriptsize{$\pm$0.001}&0.583&0.115\scriptsize{$\pm$0.001}&0.611\\
& MLP(GAT $\oplus$ Qwen3-8B) &\second{0.114\scriptsize{$\pm$0.001}}&\first{0.477}&{0.113\scriptsize{$\pm$0.001}}&0.620&0.112\scriptsize{$\pm$0.001}&0.757\\
& MLP(GAT $\oplus$ TE3L) &\first{0.112\scriptsize{$\pm$0.001}}&0.544&{0.110\scriptsize{$\pm$0.001}}&\second{0.514}&{0.111\scriptsize{$\pm$0.001}}&0.611\\
\cmidrule(l){2-8}
&GCN (Gemma3) &\second{0.114 \scriptsize{$\pm$ 0.002}} & 0.849 & 0.119 \scriptsize{$\pm$ 0.001} & 0.832 & 0.117 \scriptsize{$\pm$ 0.001} & 0.803\\
& SAGE (Gemma3) & 0.121 \scriptsize{$\pm$0.003} & 0.917 & 0.122 \scriptsize{$\pm$0.001}& 0.899 & 0.122 \scriptsize{$\pm$ 0.002} & 0.848\\
& GAT (Gemma3) & 0.116 \scriptsize{$\pm$0.001} &0.823& 0.119 \scriptsize{$\pm$0.001} & 0.849 & 0.117 \scriptsize{$\pm$ 0.001} & 0.664\\
\midrule
\multirow{2}{*}
{\rotatebox[origin=c]{90}{\small \textbf{\makecell{Temp.}}}}& \makecell{MLP($\overline{\text{GAT}} \oplus \overline{\text{TE3L}}$)}& N/A&N/A &\first{0.107\scriptsize{$\pm$0.0001}}&{0.520}&\first{0.107\scriptsize{$\pm$0.001}}&\second{0.515}\\
&\makecell{
MLP($\|\text{GAT}\|\oplus\|\text{TE3L}\|$)}& N/A&N/A &\second{0.109\scriptsize{$\pm$0.001}}&0.561&\second{0.109\scriptsize{$\pm$0.001}}&0.554\\
\bottomrule
\end{tabular}
}
}
\end{table}

\paragraph{\textbf{Datasets.}}
We present two labeled datasets per month for regression and binary classification tasks. Their label distributions are shown in Table~\ref{tab:labels_breakdown}.
\begin{itemize}[leftmargin=*]
\item \textbf{Regression Dataset:} As explained in section \ref{sec:datasets}, we use the domain quality rating dataset DQR~\cite{dataCredibility} and its extracted textual content from each \dataname dataset to train and test our regression models. We split the dataset per month into 60\%, 20\%, and 20\% for train, validation, and test set splits.

\item \textbf{Binary Classification Dataset:} As introduced in Section~\ref{sec:datasets}, we use our curated \weak classification dataset, which is 60× larger than the regression dataset and consists of approximately 662K domains labeled as credible and non-credible. These labels serve as targets for the binary classification task. For each month, the dataset is split into 60\%, 20\%, and 20\% for training, validation, and testing, respectively. Because the number of legitimate and illegitimate domains is typically imbalanced, largely due to the frequent removal of illegitimate domains, we apply a downsampling \cite{downsampling} strategy during training to mitigate class imbalance. 
\end{itemize}

\paragraph{\textbf{Credibility Prediction Baselines.}}   We compare our methods with a variety of baselines. We implemented two graph-based methods: the LightGBM model as introduced for the task in~\cite{Kadkhoda2025DomainQuality}, and the GNN trained on Search-Engine Optimization (SEO)-attributes proposed in~\cite{DBLP:conf/icwsm/CarragherWC24}. Both are supervised learning methods operating on feature-attributed graphs, relying on features from proprietary API services. Therefore, we reproduce the subset of their features that is computable from \dataname data, e.g from domain in- or out-degree. 
We also use the LLM-based prediction framework established in~\cite{DBLP:conf/websci/YangM25}, which provides an LLM with a domain name within a credibility prediction prompt. We further compare results with an augmented version of this framework, which provides the LLM with Web Search capabilities. All LLM baselines are ran on \texttt{GPT-5-mini}.



\paragraph{\textbf{ML Models.}} To examine our research questions, we implement a range of ML models across the data modalities in \dataname. 

\begin{itemize}[leftmargin=10pt]
    \item \emph{Text-Based Models:}  We use LLM models of varying parameter sizes to embed the text content into a latent vector, including:
    \textit{EmbeddingGemma-300M} (Gemma3)~\cite{EmbeddingGemma}, \textit{Qwen3 (0.6B and 8B)} embedding models \cite{qwen3technicalreport}, and \textit{OpenAI Text Embedding (TE3L)}~\cite{OpenAI-TEL3}.  The text embeddings are fed to the regressor for prediction. 
    \item \emph{Graph-Based Models:}  
    We utilize several Graph Neural Network (GNN) models: GCN\cite{GCN}, GraphSAGE\cite{SAGE}, GAT\cite{GAT}, to learn the hyperlink structure among domains and predict credibility. 
    \item \emph{GNN (Text Embedding):} 
    A GNN model where node features are initialized with their respective text embeddings, thereby enriching the structural pattern learning with text signals. 
    \item \emph{Graph $\oplus$ Text Model:}  
    We also concatenate graph embeddings (from GAT) and text embeddings (from LLMs) to form joint feature representations, which we refer to as the \emph{graph $\oplus$ text} models.  
    These concatenated embeddings are fed into an MLP regressor or binary classifier for credibility prediction. Under temporal settings (Temp.), we concatenate the averaged embeddings of the two modalities across subsequent monthly snapshots. as in  \emph{MLP($ \overline{\text{GAT} } \oplus \overline{\text{TE3L)}}$)} or concatenate them all as in    \emph{MLP($\|\text{GAT}\|\oplus\|\text{TE3L}\|$)}.
\end{itemize}

    

We report regression results for the PC1 score in Table \ref{tab:regression-all} and the binary classification accuracy in Table \ref{tab:binary-all} across all three snapshots in \dataname. To better understand models' performance in the case of regression, we report both the Mean Absolute Error~(MAE) as well as the Maximum Absolute Error~(AE$_{max}$). The MAE reflects overall performance across many domains, while the AE$_{max}$ reflects the worst possible deviation from the ground-truth score. For classification, we use traditional accuracy and F1 scores. 

\begin{table}[t]
{
\caption{Binary classification accuracy (\%) across all models. Best \first{first} and \second{second} results are highlighted. Performance is consistent across months, with 3-month aggregation performing best. An asterisk denotes our added web search.}

\label{tab:binary-all}
\centering
\setlength{\tabcolsep}{4pt} %
\resizebox{0.48\textwidth}{!}{%
\begin{tabular}{ll l l l}
\toprule
& \textbf{Method} & \textbf{CDB-Oct24} & \textbf{CDB-Nov24} & \textbf{CDB-Dec24} \\

\midrule

\multirow{5}{*}{\rotatebox[origin=c]{90}{\small \textbf{Baselines}}} &
 Constant &52.2&58.9&53.0 \\ 

& SEO-based GNN~\cite{DBLP:conf/icwsm/CarragherWC24} & 50.0 & 50.0 & 50.0 \\ 
& LightGBM~\cite{Kadkhoda2025DomainQuality} & 56.0 & 56.0 & 56.0 \\ 
& LLM-URL~\cite{DBLP:conf/websci/YangM25} & 52.8 & 52.8 & 52.8 \\
& LLM-URL~\cite{DBLP:conf/websci/YangM25} + WS* &
54.35 & 54.35 & 54.35 \\ 

\midrule

\multirow{4}{*}{\rotatebox[origin=c]{90}{\small \textbf{Text}}} & MLP(Gemma3 (256)) &71.9\scriptsize{$\pm$0.05}&73.5\scriptsize{$\pm$0.04}&72.8\scriptsize{$\pm$0.002}\\
& MLP(Qwen3-0.6B(256)) &69.5\scriptsize{$\pm$0.06}&70.6\scriptsize{$\pm$0.01}&70.4\scriptsize{$\pm$0.01}\\
& MLP(Qwen3-8B(1024))  &70.1\scriptsize{$\pm$0.01}&69.8\scriptsize{$\pm$0.03}&69.9\scriptsize{$\pm$0.06}\\
& MLP(TE3L(3072))  &61.4\scriptsize{$\pm$0.02}&64.5\scriptsize{$\pm$0.01}&61.4\scriptsize{$\pm$0.02}\\
\midrule

\multirow{3}{*}{\rotatebox[origin=c]{90}{\small \textbf{Graph}}} &
GCN &67.4\scriptsize{$\pm$0.27}&63.2\scriptsize{$\pm$0.59}&68.0\scriptsize{$\pm$0.68}\\
& GraphSAGE  &67.2\scriptsize{$\pm$0.6}&62.1\scriptsize{$\pm$0.41}&66.8\scriptsize{$\pm$0.79}\\
& GAT &69.8\scriptsize{$\pm$0.53}&65.4\scriptsize{$\pm$0.37}&68.9\scriptsize{$\pm$0.18}\\
\midrule
\multirow{7}{*}{\rotatebox[origin=c]{90}{\small \textbf{\makecell{Text \& Graph}}}}& MLP (GAT $\oplus$ Gemma3(256)) &\first{83.4\scriptsize{$\pm$0.02}}&{83.7\scriptsize{$\pm$0.04}}&{83.4\scriptsize{$\pm$0.03}}\\
& MLP (GAT $\oplus$ Qwen3-0.6B(256))& \second{83.3\scriptsize{$\pm$0.05}}&\second{84.1\scriptsize{$\pm$0.02}}&83.6\scriptsize{$\pm$0.03}\\
\cmidrule(l){2-5}
& GCN (Gemma3) & 75.7\scriptsize{$\pm$0.73}&73.1 \scriptsize{$\pm$0.84}&76.5 \scriptsize{$\pm$0.18} \\
& SAGE (Gemma3) & 74.1\scriptsize{$\pm$1.14}&72.0\scriptsize{$\pm$0.64}&74.5 \scriptsize{$\pm$0.2}\\
& GAT (Gemma3) &75.7\scriptsize{$\pm$0.47}&74.4\scriptsize{$\pm$1.08}&76.1 \scriptsize{$\pm$0.36}\\
& GAT (Gemma3) &75.7\scriptsize{$\pm$0.47}&74.4\scriptsize{$\pm$1.08}&76.1 \scriptsize{$\pm$0.36}\\
\midrule
\multirow{2}{*}{\rotatebox[origin=c]{90}{\small \textbf{\makecell{Temp.}}}}
& MLP($\overline{\text{GAT}}\oplus\overline{\text{Gemma3}}$) &N/A&83.9\scriptsize{$\pm$0.01}&\second{84.1\scriptsize{$\pm$0.03 }}\\
& MLP($\overline{\text{GAT}}\oplus\overline{\text{Qwen3-0.6B}}$) &N/A&\first{84.6\scriptsize{$\pm$0.05}}&\first{85\scriptsize{$\pm$0.01}}\\
\bottomrule
\end{tabular}
}
}
\end{table}

\subsection{Experimental Results}
\paragraph{\textbf{RQ1: Credibility prediction with text content.}}
Here, we aim to answer if the extracted web text content contains beneficial signals for predicting domain credibility. This question is reflected by the performance of \emph{text-based} models as shown in Table \ref{tab:dec-all}. All models using learned text features outperform the baselines. Figure \ref{fig:mlp-pc1-dist} further illustrates how this model, under regression, learns the credibility distribution scores. This confirms that the text content in \dataname provides significant signals for credibility prediction when embedded. Indeed, we find that \emph{text-based} models outperform standalone graph usage. This observation is consistent across other snapshots as seen in Table \ref{tab:regression-all} and Table \ref{tab:binary-all}. Under credibility regression scores, we observe that the LLM with the most parameters, i.e. OpenAI-TE3L, is the best performer in the text-only category. In addition, within the same Qwen3 family, more parameters also enable better performance.  Moreover, the augmented LLM baseline, which has Web Search enabled, provides more insightful results than its native version. 

\vskip .1in

Exclusively LLM-based assessments seem insufficient, attaining a MAE of around 15\% with a Max AE of above 70\%. To further analyze the reliability of LLMs as source credibility predictors, we look at their refusal rates; that is, the number of domains for which the model refuses to provide a rating. This happens almost exclusively when the LLM lacks sufficient information on the domain, to provide any prediction. We report the quantity of such rates in Appendix~\ref{app:refusal}; most notably, we observe substantially higher refusal rates in models not equipped with Web Search. Indeed, between 20\% (for regression) and 40\% (for classification) of domains are refused, whereas this percentage decreases to about 5\% with Web Search enabled. Such refusal rates demonstrate the insufficiency of non-augmented LLMs for the task.

\paragraph{\textbf{RQ2: Credibility prediction with graph structure.}} To understand the impact of graph structure in modeling credibility, we train GNNs on our three snapshots in \dataname. For all GNN models, we started with the Random Node Initialization~\cite{RNI} (RNI) as the node feature, normally distributed in $\mathcal{N}(0, 1)$.
In Table ~\ref{tab:dec-all}, we observe that GNN models consistently outperform the baseline across each snapshot as seen in Tables  \ref{tab:regression-all} and \ref{tab:binary-all}. 
In particular, we see that the GAT model is the best-performing graph model. Figures \ref{fig:gat-rni-pc1-dec} further illustrate that the GAT model learns the distribution of credibility scores. Indeed, this indicates that the hyperlink graph structure provides a useful signal for credibility prediction. Notably, in the Max AE metric, graph-based methods have significantly less performance when compared to text-based models. Figure \ref{fig:gat-rni-pc1-dec} and \ref{fig:mlp-pc1-dist} show that while GAT learns the distribution, it deviates more from the target distribution compared to the text-based MLP models. This might be caused by low-degree domains or a lack of hyperlinks on some domains.

\paragraph{\textbf{GNN Neighbor Sampling Ablation.}} 
To ablate the graph learning models, we report the effects of varying the number of neighbours and hops in sampling. We report Out of Memory as OOM. Here, generally as more number of neighbors are sampled, the GNN MAE performance on the PC1 score has improved. Similarly, in most cases, increasing the number of hops also improves the MAE score. Note that here, we also use the RNI node features same as in RQ2. Interestingly, unlike link prediction or node classification~\cite{DBLP:conf/nips/0004S0L23, DBLP:conf/iclr/XuRKKA20}, which heavily depends on one or two hops, the credibility assignment task achieves best performance at 3 hop (with 30 neighbours each) on the GAT model. This suggests that higher-order neighbour information is crucial for the credibility assignment task.

\begin{table}[t]
\caption{Ablation study of GAT's MAE for PC1 score regression with varying numbers of neighbours sampled and hops. Large-scale one-hop settings are shown separately.}
\centering
\begin{tabular}{lccc}
\toprule
\textbf{Num Neighbors} & \multicolumn{3}{c}{\textbf{PC1 (MAE)}} \\
\midrule
 & \textbf{1-hop} & \textbf{2-hop} & \textbf{3-hop} \\
\midrule
10{,}000   & 0.144 \scriptsize{$\pm$ 0.001} & N/A & N/A \\
125{,}000  & 0.145 \scriptsize{$\pm$ 0.001} & N/A & N/A \\
\midrule
5   & 0.147 \scriptsize{$\pm$ 0.001} & 0.148 \scriptsize{$\pm$ 0.001} & 0.150 \scriptsize{$\pm$ 0.002} \\
10  & 0.141 \scriptsize{$\pm$ 0.004} & 0.143 \scriptsize{$\pm$ 0.003} & 0.138 \scriptsize{$\pm$ 0.002} \\
30  & 0.138 \scriptsize{$\pm$ 0.003} & 0.137 \scriptsize{$\pm$ 0.001} & \textbf{0.129 \scriptsize{$\pm$ 0.002}} \\
50  & 0.142 \scriptsize{$\pm$ 0.002} & 0.135 \scriptsize{$\pm$ 0.001} & OOM\\
\bottomrule
\end{tabular}
\label{tab:gat-ablation}
\end{table}

\paragraph{\textbf{RQ3: Combining graph and text.}}
\dataname contain rich graph and text features; as such we believe that leveraging these modalities in conjunction may provide the best performance in credibility prediction. Indeed, this hypothesis is confirmed emprically through the our \textit{GNN (Text)} and \textit{Graph $\oplus$ Text} models. In Table \ref{tab:dec-all} we observe that the best performing model is \textit{Graph $\oplus$ Text}, which scores significantly better than all baselines and other models. Similarly, the \textit{GNN (Text)} model, which utilizes text features, has improved performance compared to graph-only. This observation is consistent across all snapshots in a regression task in Table \ref{tab:regression-all} and a binary classification task in Table \ref{tab:binary-all}. To assess the stability of our binary classifier, we evaluated its performance using multiple metrics.  MLP(GAT) consistently achieved the best performance, with a Recall of 83.4\%, AUROC of 83.4\%, and AUPRC of 79.5\%. Similar performance trends were observed across all evaluated models. All model results will be provided in the dataset card. Additionally,  we observe that the distributions of the \textit{Graph $\oplus$ Text} models in Figure \ref{fig:regression-dist-dec-comparison} align closely with the target distribution. These results support that a multi-modal setting is superior for credibility prediction.

\paragraph{\textbf{RQ4: Content and Graph Structure Evolution}}A simple approach to capture graph and content evolution over three months is to aggregate and then concatenate textual and structural embeddings per month, then feed them to an MLP regressor or classifier. As shown in Tables~\ref{tab:regression-all} and \ref{tab:binary-all}, temporal graph $\oplus$ text models achieve the best performance. For CDB-Dec24 we average three-month embeddings, while for CDB-Nov24 we aggregate only October and November. On CDB-Dec24, the binary classification accuracy improves from 83.6\% to 85\%, while the regression MAE decreases from 0.111 to 0.107. Incorporating additional temporal information further improves the accuracy from 84.6\% in November to 85\% in December and reduces the PC1 Max(AE) from 0.520 to 0.515. These results underscore the importance of modeling the temporal evolution and edges existence timestamps for credibility scoring and motivate future temporal GNN $\oplus$ LLM models on our \dataname.

\begin{table}
\centering
\caption{Cross-domain transfer performance of MLP(GAT $\oplus$ Gemma3) on CDB-Dec24. Each model is trained on a single domain type and evaluated on the full test set (all domain types), measuring how well domain-specific knowledge transfers to unseen domain types using ACC (\%) and F1 scores. The \first{first} and \second{second} results are highlighted.}
\resizebox{0.45\linewidth}{!}{%
\begin{tabular}{ccc}
\toprule
\textbf{Category} & \textbf{ACC (\%)} & \textbf{F1} \\
\midrule
All Domains & \first{77.4} & \first{77.3} \\
Phishing    & \second{76.4} & \second{76.2} \\
Mis-Info    & 47.1 & 32.0 \\
General     & 66.0 & 64.9 \\
\bottomrule
\end{tabular}}
\label{tbl_transferability}
\end{table}

\paragraph{\textbf{RQ5: Transferability Across Domains; different notions of credibility.}}
Since the DomainRel dataset is heavily skewed toward phishing domains, we analyze its impact on model performance and generalization across other misinformation types. Table~\ref{tbl_transferability} reports model transferability when trained on individual domain categories and evaluated on the full test set. Training on all domains yields the strongest overall performance. Despite having fewer labels than phishing, the general-domain subset also achieves competitive results. We exclude the malware domains subset as it contains only non-credible labels, resulting in severe class imbalance.

\paragraph{\textbf{RQ6: Transferability Across Time.}}
\begin{table}[!t]
\caption{Month-to-month temporal transfer. A model trained on a given month is evaluated on the next month. The \first{first} and \second{second} best ACC(\%) are highlighted.}
\begin{tabular}{llll}
\toprule
\textbf{Method} & \textbf{CBD-Oct24} & \textbf{CBD-Nov24} & \textbf{CBD-Dec24} \\ \midrule
MLP (Oct, Gemma3) & 71.9 & 73.3 & - \\
GAT (Oct, Gemma3) & \first{81.4} & \second{81.2} & - \\
GAT (Oct, RNI) & \second{73.0} & 47.5 & - \\
\midrule
MLP (Nov, Gemma3) & - & 73.5 & \second{71.6} \\
GAT (Nov, Gemma3) & - & \first{81.5} & \first{79.2} \\
GAT (Nov, RNI) & - & 74.3 & 54.0 \\ \bottomrule
\end{tabular}
\label{tbl:time-transferability}
\vspace{-1em}
\end{table}

We evaluate temporal transferability by measuring how models trained on one monthly snapshot generalize to the subsequent month. As shown in Table~\ref{tbl:time-transferability}, GAT(RNI) exhibits poor transferability compared to models leveraging text features. This is likely because the web graph structure changes substantially across months (Table~\ref{tab:transition_stats}), limiting the effectiveness of structure-only representations. 


In contrast, GAT(Gemma3), which jointly incorporates graph and textual features, achieves the strongest temporal transferability, while models trained directly on the inference month attain the highest overall performance. We also observe that each monthly snapshot introduces a large proportion of domains with previously unseen textual content. The table below reports the ratio of new text content among labeled domains. Together with the transferability results, this suggests that the models learn meaningful temporal generalization from textual features rather than simply memorizing domain content.

\section{Conclusion}

In this work, we introduced an optimized web graph construction pipeline, accompanied by the large \dataname dataset. On these unprecedentedly large graph datasets, we explore the strength of different signals, including structural and text content, in predicting credibility. Our empirical findings show that it is important to model all modalities of web domains including the internet topology, temporality and web page text content in order to achieve better credibility score assignment. In addition, \dataname datasets are text-attributed temporal graphs thus they serves as a strong starting point for future work to explore more aspects in the temporal evolution of web domains and gain better understanding on how to combat misinformation. Through this work, we hope to facilitate the development of scalable, ML-based misinformation detection systems that enhance information veracity to build more trustworthy online information ecosystems.

\begin{acks}
This research was supported by the Engineering and Physical Sciences Research Council (EPSRC) and the AI Security Institute (AISI) grant: Towards Trustworthy AI Agents for Information Veracity and the EPSRC Turing AI World-Leading Research Fellowship No. EP/X040062/1 and EPSRC AI Hub No. EP/Y028872/1. This research
was also enabled in part by CIFAR and IVADO. The computing resources are provided by Mila (Quebec) and Compute Canada.
\end{acks}


\bibliographystyle{ACM-Reference-Format}
\bibliography{references}


\begin{table}[!h]
\centering
\caption{Source Composition of DomainRel, 
with counts for Reliable (R) and Unreliable (UR) domains.}
\label{tab:labels_breakdown_indetail}
\setlength{\tabcolsep}{5pt}
\resizebox{0.48\textwidth}{!}{
\begin{tabular}{llrrr}
\toprule
\textbf{Domain} & \textbf{Total} & \textbf{Dataset} & \textbf{UR(0)} & \textbf{R(1)} \\
\midrule

\multirow{1}{*}{\makecell{\text{Crowd}}}
& 3{,}935 &  Wikipedia~\cite{kynoptic_wikipedia_reliable_sources_2026}
& 1{,}029 & 2{,}906 \\
\midrule

\multirow{2}{*}{\makecell{MisInfo}} & \multirow{2}{*}{4,818} 
& Nelež~\cite{nelez_disinformation_sites_dataset_2025}
& 51 & 0  \\
& & Misinfo Domains~\cite{lasser_misinformation_domains}
& 2{,}170 & 2{,}597 \\
\midrule

\multirow{4}{*}{Phishing} & \multirow{4}{*}{649,253} &
LegitPhish~\cite{LegitPhish}
& 26,871 & 37{,}113 \\
& & PhishDataset~\cite{PhishDataset}
& 3{,}730 & 40{,}535 \\
& & URL-Phish~\cite{damminh_urlphish_2025}
& 10{,}551 & 92{,}460 \\
& & Phish \& Legit~\cite{phish_and_legit}
& 291,742 & 146,250 \\
\midrule

\multirow{1}{*}{Malware}
& 31{,}540 &  Malware Domains~\cite{URLhaus}
& 31{,}540 & 0 \\
\midrule

\textbf{Total}
& \textbf{662,557}& & \textbf{356,403} & \textbf{306,172} \\
\bottomrule
\end{tabular}}
\end{table}


\appendix

\begin{table}[!t]
\caption{Time breakdown of the pipeline steps per 1 batch in Minutes (where one month = 300 batches)}
\label{tab:time-steps}
\centering
\resizebox{0.48\textwidth}{!}{%
\begin{tabular}{cccc}
\toprule \textbf{Step} &\textbf{Data Download} & \textbf{Processing} & \textbf{Aggregation} \\
\midrule
Graph Construction&10  & 210  & 10  \\ 
Index Filtering&0.1  & .8  & 0.1  \\ 
Content Extraction&2.5 &5& 0.5  \\ 
Content Embedding&N/A&10&N/A  \\ 
\bottomrule 
\end{tabular}}
\end{table}

\section{Text-Attributed temporal Graph Construction Pipeline}
\label{app:CG}

We designed an efficient graph construction pipeline optimized for performance, resource usage, and usability. Below, we describe the data format, pipeline, and processing protocols.

\paragraph{Data Format.}
The data provided by Common Crawl is in Web ARChive (WARC) files \cite{WARC_format}, the industry standard for web archiving. Metadata is found in Web ARChive Timestamp (WAT) files, containing HTTP response headers, links extracted from HTML pages and other metadata. Additionally, WARC Encapsulated Text (WET) files contain extracted plain text from web content.

\paragraph{TAG Construction.}
Raw link data is obtained from Common Crawl monthly Web Graphs ~\cite{Common-Crawl} and processed in distributed (multi-CPU, we use a default of 16) bash scripts to build the graph. This includes building the structure itself (for each monthly snapshot, vertices in \texttt{<domain>, <timestamp>} format and edges in \texttt{<src>, <dst>, <timestamp>}) format. This is following 
by graph processing, where nodes are deduplicated and ones of degree $< 3$ removed, and domain names are normalized. Text extraction is then done for all the domains present in the processed graph: their webpage content is found either in Common Crawl raw data, or scraped directly on the web. It is then embedded. 


\paragraph{Compute \& Resource Optimization.}

For the data downloading, decompressing and graph construction, constructing one monthly snapshot takes around 120 hours for all batches with 128GB of RAM and 16 CPU cores. Downstream degree-based filtering and target label generation for supervised learning consume an additional 7 hours per monthly snapshot of upwards of a billion edges, with the same computational setup. 
To enable this, this step offloads sorting and joins to disk-backed external processed and temporary files to avoid costly in-python accumulators. We also use memory-mapped degree vectors to avoid loading large structures into RAM, and performing streaming, sequential passes that minimize in-memory state. The result pipeline processes massive graphs efficiently while remaining resource-bounded and reproducible. Table~\ref{tab:time-steps} details the cost breakdown of processing per batch. 

\paragraph{Content Extraction and Embedding Pipeline.}
\label{appendix:content_extract}
Our optimized pipeline for extracting representative textual content from each monthly graph snapshot generates embeddings from web crawls for downstream tasks. It leverages the Common Crawl monthly snapshot index to avoid scanning and downloading roughly 50\% of the data, which amounts to approximately 7TB per month. The pipeline extracts textual content from snapshot WET files and stores it in an indexed, columnar Parquet format, enabling fast and parallelized LLM embedding generation. To maintain representative content samples per domain, we retain six documents of varying content length for each domain. For labelled domains lacking textual content, we employ a multi-threaded online scraping pipeline in batches to extract text directly from the domain’s home page.
At this scale, content extraction and embedding are typically highly resource-intensive, taking several weeks by default. Our fully parallelized pipeline reduces content extraction to 1.5 day and embedding generation to approximately 2 days, depending on the embedding model size, facilitating large-scale content representation suitable for graph-based and retrieval-augmented applications. Index filtering and text extraction jobs run as parallel SLURM jobs on CPU machines, each equipped with 64 GB of RAM and 64 vCPU cores. The content embeddings were generated separately on 12 machines, each equipped with NVIDIA L40S GPU, 32 GB of RAM, and 4 CPU cores.

\paragraph{Text embedding generation.} The content of each domain is transformed into an embedding using the Qwen3-0.6B multilingual embedding model \cite{qwen3technicalreport}, which produces 1024-dimensional representations per domain that could be down-sized into smaller dimensions (i.e., 128) as it supports Matryoshka Representation Learning (MRL) \cite{MRL}. Notably, Qwen3-0.6B ranks fourth on the MTEB leaderboard while being the smallest model among the top-performing models \cite{MTEB-Leaderboard}. Figure \ref{fig:domains_evolution}.c  illustrates the relative sizes of domains with textual content across the three monthly snapshots.
Moreover, ~50\% of the DQR domains and ~60\% of the DomainRel domains have content across the 3 months, as shown in the Huggingface dataset card.
\paragraph{Graph sampling for Compute-Efficient GNN Training.} To facilitate graph learning on our large-scale graph, we use a neighbour loader sampler (sampling multi-hop neighbours) to sample adequate training batches of size $1024$ at 3-hops ($[30, 30, 30]$). Similarly, to handle large-scale feature matrices, we reduced the dimensions of node embeddings to $64$. The experiments were run using an 80GB memory NVIDIA A100 Tensor Core GPU.

\section{Experimental Setup}
\label{app:exps}

\paragraph{The DQR dataset splitting.} Domains are stratified according to the PC1 score and partitioned into training, validation, and test sets comprising 60\%, 20\%, and 20\% of the data, respectively.

\paragraph{MLP Experiment.}  A scikit-learn MLP regressor with two hidden layers (the first of dimension 128, the second 64) with a ReLU activation function, an Adam optimizer, and a learning rate of 0.001 is applied. The models were trained for 15 epochs with a maximum number of iterations set to 200.

\paragraph{Domain content examples with their PC1.} The following examples present text content extracted from CC-WET files for various DQR domains, along with their corresponding credibility scores. The higher the score, the more credible the domain.

\paragraph{GNN Experiment.} 
We used different GNN architectures, such as residual connections, which performed well in node-regression tasks \cite{graphland}. Each GNN has $3$ layers, a dropout of $0.1$, and a learning rate of $0.001$ on the Adam optimizer. Due to the scale of our dataset, we employed the use of PyG's NeighborLoader \cite{DBLP:journals/corr/abs-1903-02428, DBLP:journals/corr/abs-2507-16991}, a neighbor sampling method introduced in \cite{SAGE} to mini-batch our data. Each GNN runs for 100 epochs over 3 trials.

\section{LLM Refusal Rates}
\label{app:refusal}

Table \ref{tab:refusal-rates} shows the rates of queries that LLMs refuse to answer, with or without Web Search enabled.
These highlight the limitations that LLMs present out-of-the-box, and the brittleness that remains in retrieval processes; domains can still disappear, be renamed, or blocked by paywalls, and retrieval pipelines are still underdeveloped.

\begin{table}[h]
\caption{When web search is not activated, LLMs frequently lack the information needed to answer a domain query. When close to a quarter of queries get refused this way, this makes off-the-shelf LLMs unreliable for the task.}
\label{tab:refusal-rates}
\centering
\begin{tabular}{l cc c}
\toprule
\textbf{Configuration} & \textbf{N.Queries} & \textbf{N.Refusals} & \textbf{Refusal rate} \\
\midrule
Base Model & 1722  & 330 & 19.15\% \\
Base + Web Search & 1722 & 98 & 5.6\%   \\
\bottomrule
\end{tabular}
\end{table}

\end{document}